\documentclass[preprint,preprintnumbers,amsmath,amssymb]{revtex4}
\usepackage{graphicx}
\usepackage{dcolumn}
\usepackage{bm}

\begin{document}

\title{Scattering from rough surfaces: A simple reflection phenomenon in fractional space}
\author{H. Safdari $^1$, M. Vahabi $^2$, G.~R. Jafari $^1$ \\
{\small $^1$ Department of Physics, Shahid Beheshti University,
G.C., Evin, Tehran 19839, Iran} \\
{\small $^2$ Labratoire de Physique de la Mati\`{e}re Condens\'{e}e
(UMR 7643), CNRS-Ecole Polytechnique, F-91128, France}}
\date{\today}


\begin{abstract}

In this paper, scattering of incident plane waves from rough
surfaces have been modeled in a fractional space. It is shown how wave
scattering from a rough surface, could be a simple reflection problem in a
fractional space. In the integer space, fluctuations of the surface
result in wave scattering while in the fractional space these
fluctuations are compensated by the geometry of the space. In the
fractional space, reflection leads to the same results as the
scattering in the integer space. To make it more clear, scattered
wave function in the framework of Kirchhoff theory is considered in
a fractional space and results are compared with those from a
self-affine surfaces. Our results show that these two approaches are
comparable.

\end{abstract}


\maketitle

\section{Introduction}

The problem of wave scattering from rough surfaces has been of
substantial interest for more than a century. Extensive studies have
been carried out at experimental and theoretical levels until now
\cite{Germer,Elfouhaily,Salami}. Indeed, this problem has
gained considerable attention in diverse areas of science and
engineering \cite{Schröder,Xuab,Khenchaf,Sinha}. In reality, no surface is ideally
smooth. In fact, wave scattering from a surface is affected by the
morphology and roughness of this surface. Small perturbation method
(SPM) \cite{Johnson} and Kirchhoff theory
\cite{Ogilvy,Voronovich,Jafari} are the two oldest and most
employed approximate approaches to this problem.

When the surface under study is slightly rough, then SPM can be used
for finding the solution to the scattering problem. In this case,
the rough surface is assumed as a height perturbation to a smooth
surface and the resulting changes due to the roughness is considered
in the scattering coefficient. This approach requires that the
height function is everywhere small compared to the wavelength of
the incident wave and its gradient is also small in comparison to
unity \cite{Ogilvy} but it is independent from the radius of
curvature of the surface. When the points on the surface have a
large radius of curvature relative to the wavelength of the incident
field and the surface roughness may not be small compared to it,
Kirchhoff theory can be applied. Kirchhoff theory, also known as the
"tangent plane theory" is used in conjunction with an integral
formula, to give an expression for the scattered field from the surface in terms of the approximated surface
field.  The physical basis for this theory is tangent plane
approximation: any point on the surface is assumed to have the same
optical behavior as if the surface was locally flat. In fact, the
scattered field and its normal derivative at the boundary can be
expressed through the incident field which help to reconstruct the
scattered field in the total space \cite{Voronovich}. Here, we focus
on the Kirchhoff theory which is a local approximation method and
could be used for surfaces much rougher than those considered by SPM
\cite{Ogilvy}. We use the approach of the fractional calculus to
deal with this scattering problem. In general, fractional calculus
is the generalization of the classical calculus which deals with the
integrals and derivatives of arbitrary (real or even complex) orders
\cite{MILLER,Igor,Metzler}. Remarkable attention and significance for this
area have been achieved in recent decades because of its diverse
application in varies fields of science and technology ranging from
physics \cite{Barkai,Tarasov,Wilczek,Nonnenmacher}, plasma \cite{Vahabi} and polymer \cite{HEYMAN}
to engineering \cite{Sabatier,Carpinteri}, biology
\cite{Magin,Bronstein,Brockmann}, finance \cite{Mainardi} and ... .
Indeed, there are situations that fractional calculus are so useful.
For instance, the concept of nonlocality, memory and hereditary
properties could be imported using the fractional operators.

Here, we consider the problem of scattering of a monochromatic plane
wave from a rough surface and reformulate the problem in a
fractional space. In the latter space, which is described by a
constant non-integer fractal dimension, the surface is not rough
anymore.  In the integer space, fluctuations of the surface lead to
wave scattering while in the fractional space these fluctuations are
compensated by the topology of the space. Thus, the problem changes
into the reflection of plane waves from a flat surface in a
fractional space of order $\alpha$, $2<\alpha<3$.  By considering
the scattered wave function in the framework of Kirchhoff theory we
show how reflection from a flat surface in the fractional space
leads to the same results as the scattering in the integer space for
the self-affine surfaces.

The paper is organized as follows. In Sec. II, we briefly review the
Kirchhoff theory of wave scattering. In Sec. III, we discuss our
method and our results are presented. Finally, Sec. IV summarizes
our conclusion.

\section{Kirchhoff theory of wave scattering}

\begin{figure}[t]
\includegraphics[width=5.5cm,height=5.5cm,angle=0]{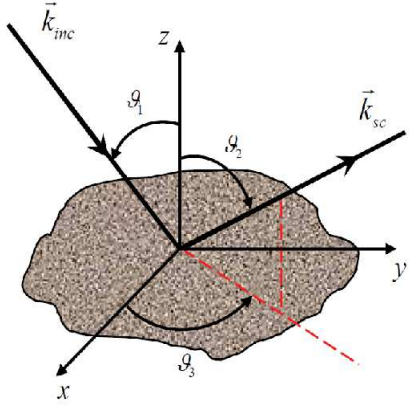}
\caption{The geometry used to investigate wave scattering from a
rough surface.}\label{Fig1}
\end{figure}

In Kirchhoff theory of wave scattering, valid in the far field
region of the rough surface, we consider a
incident plane wave, $\psi^{inc}(r)=\mathrm{exp}(-i k_{inc}.r)$,
$k$ and $r$ stand for wave number and position, respectively. The
geometry used to study the scattering phenomena from a rough
surface, is shown in Fig. \ref{Fig1}. By making the assumption that
there is no point on the scatterer surface with infinite gradient,
we work under the Dirichlet boundary condition, i.e., the surface
reflectance, $R_{0}=-1$. Therefore, the total scattered field,
$\psi^{sc}(r)$, over the mean reference plane $A_{M}$, is given by
\cite{Ogilvy}
\begin{eqnarray}
\psi^{sc}(r) &= & \frac{ike^{ikr}}{4\pi r}\int_{A_{M}}
\left(a\frac{\partial h}{\partial x_{0}}+b\frac{\partial h}{\partial
y_{0}}-c\right) \nonumber \\ \nonumber\\
& & \times\mathrm{exp}\left(ik[Ax_{0}+By_{0}+
Ch(x_{0},y_{0})]\right)dx_{0} dy_{0}, \nonumber \\
\end{eqnarray}
where
\begin{eqnarray}
A&=&\sin \theta_{1}-\sin \theta_{2} \cos \theta_{3}, \nonumber \\
B&=&-\sin \theta_{2} \sin \theta_{3}, \nonumber \\
C&=&-(\cos \theta_{1} +\cos \theta_{2}), \nonumber \\
a&=&\sin \theta_{1}(1-R_{0})+\sin \theta_{2}\cos \theta_{3}(1+R_{0}), \nonumber \\
b&=&\sin \theta_{2} \sin \theta_{3}(1+R_{0}), \nonumber \\
c&=&\cos \theta_{2}(1+R_{0})-\cos \theta_{1}(1-R_{0}).
\end{eqnarray}

Here, $h(x_{0},y_{0})$ is the height of the surface at position
$(x_{0},y_{0})$ from the reference surface. The total scattered
field intensity, the experimentally measurable quantity,
$I_{tot}=I_{coh}+I_{d}= <\psi^{sc}(r)\psi^{sc\ast}(r)>$, is
consisted of two parts; the coherent intensity ($I_{coh}$) and the
diffuse one ($I_{d}$). The major contribution of the coherent part
is seen in the specular direction while for the diffuse part it is
in other directions. Root mean square, $\sigma$, is a scale for the
surface roughness that affects the specular part; besides $\sigma$,
correlation function is another parameter which has only impacts on
the diffuse part. In previous works on Kirchhoff theory, these two
parts of intensity (coherent, $I_{coh}$, and diffuse, $I_{d}$,
parts) were separately studied, but here, we consider the total
scattered intensity, $I_{tot}$, itself.

\section{Equivalence of 3D scattering and reflection in the fractional space}

In our new perspective, to solve this integral and get the scattered
field, instead of dealing with roughness, we take the scatterer
surface as a completely smooth one and compensate the outcomes by
considering a fractal dimension for our space. As a consequence, the
dimension of the space is no longer the integer dimension of the
Euclidian embedding space. In other words, we have a fractional
integration of a plane wave over a flat surface. Subsequently, the
concept of fractional calculus will appear and in lieu of scattering
from a rough surface we have just reflection from a flat surface in
fractional space. The coherent intensity corresponds to reflection
in specular direction. Also there are reflections in other
directions than specular one, which are equivalent to diffuse
scattering from rough surfaces.

Since there is no height fluctuation, the mean height
from the reference surface is zero everywhere, $h(x_{0},y_{0})=0$.
Finally, the scattered field will be of the following form,
\begin{eqnarray}
\psi^{sc}(r) & &= \frac{i(-c)ke^{ikr}}{4\pi r}\int_{s_{0}}
 \mathrm{exp}\left(ik As\right)d^{\alpha}s,
\end{eqnarray}
where the integral is on the light spot size, $s_{0}$. Here $\alpha$
is the order of fractional integration which shows the fractal
dimension of the surface, in this case $2< \alpha< 3$ and $s$ is the
flat surface in fractional space. Fractal dimension or equivalently
Hurst exponent, $H=d-\alpha$ (where $d=3$ is the dimension of the
embedded space so $0<H<1$) is a measure for the roughness of the
fractal surface. For integer spaces with fluctuations, small $H$
values represent high irregularity, while for $H$ close to $1$ the
surface is more regular; hence, for fractional dimensional spaces,
small $H$ values correspond to surfaces with higher dimensionality
while large $H$ values correspond to smaller dimensions.

All we have to do now is to find the fractional integral of order
$\alpha$ of a plane wave. In fractional calculus, the $\alpha$ order
integral of an exponential function is given as \cite{MILLER},
\begin{eqnarray}
_{0}D^{-\alpha}_{x} e^{m x} = x^{\alpha}  e^{m x}
\gamma^{\star}(\alpha,m x)=E_{x}(\alpha, m),
\end{eqnarray}
where $_{0}D^{-\alpha}_{x}$ is the left Riemann-Liouville integral
of order $\alpha$, $m$ is an arbitrary constant that is complex
here, $x$ is the integral variable. In this formula,
$\gamma^{\star}$ is the incomplete Gamma function which is defined
as,
\begin{eqnarray}
\gamma^{\star}(\alpha,m x) =\frac{1}{(m x)^{\alpha}
\Gamma({\alpha})}\times \int_{0}^{m x}{\xi^{\alpha-1} e^{-\xi}}
d\xi,
\end{eqnarray}
and $E_{x}(\alpha, m)$ is the Miller-Ross function which is related
to the Mittag-Leffler functions as follows,
\begin{eqnarray}
E_{x}(\alpha, m)=x^{\alpha}E_{1,\alpha+1}(m x).
\end{eqnarray}

\begin{figure}[t]
\includegraphics[width=8.5cm,height=6.5cm,angle=0]{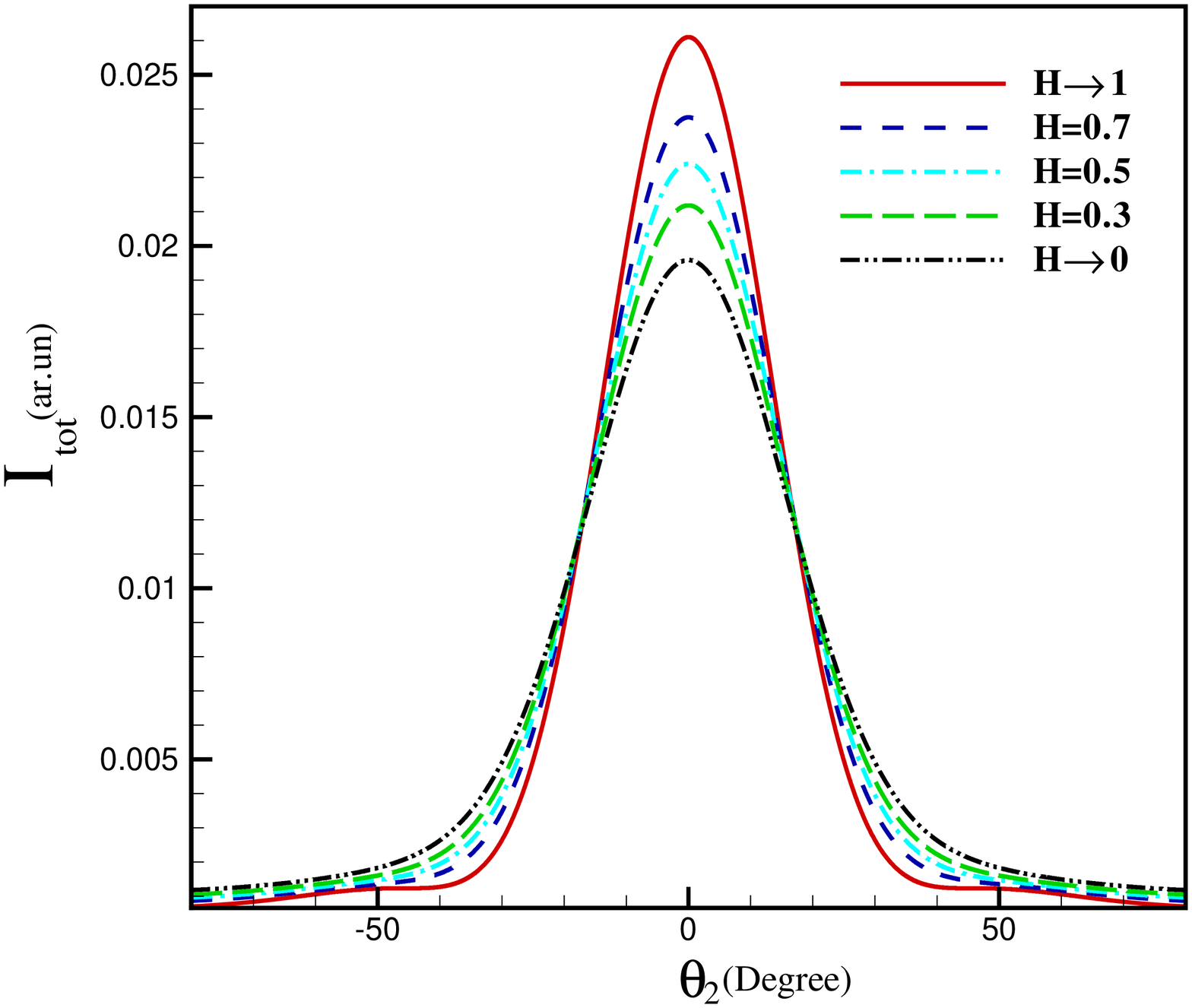}
\caption{(a) Dependence of the total scattered field intensity,
$I_{tot}$, on scattering angle, $\theta_{2}$, for different values
of Hurst exponent, vertical incident, $\theta_{1}=0$, and for
monochromatic light $\lambda=500$ $nm$ and
$\theta_{3}=0$.}\label{Fig2}
\end{figure}

Fig. \ref{Fig2} shows the dependence of the total normalized
scattered field intensity, $I_{tot}$, on the scattering angle,
$\theta_{2}$, for different values of Hurst exponent, $H$. By
decreasing $H$, fractional dimension of the surface will increase
which is equivalent to increasing the roughness of the surface in
the Euclidian space. As a result, the amount of reflected intensity
in specular direction will decrease and contribution of reflection
in other directions which appear in this approach and corresponds to
diffuse component from rough surface will increase, which is in good
agreement with the results obtained by Kirchhoff theory for rough
surfaces \cite{Zamani}.

Reflection in specular angle, $\theta_{1}=\theta_{2}$ is equivalent
to coherent part of scattering, and in other directions is its
diffuse part. So, intensity in angles far away from specular
directions, gives us the diffuse intensity.

Wavelength is the observation scale of the surface. When wavelength
of the incident beam is shorter than the correlation length of the
surface, the more the wavelength is decreased, the more the
roughness of the surface is being sensed by the incident beam. For a
flat surface in the fractional space, the correlation length is
infinite so it is larger than the incident wavelength and roughness
could be sensed more. Thus, the diffuse reflected intensity should
increase by decreasing the wavelength which could be seen in Fig. 3.
The total intensity for angles far from zero (non-specular angles)
shows the diffuse intensity and it decreases by decreasing the
wavelength which is in good agreement with the results obtained in
\cite{Zamani}.
\begin{figure}[t]
\includegraphics[width=8cm,height=6cm,angle=0]{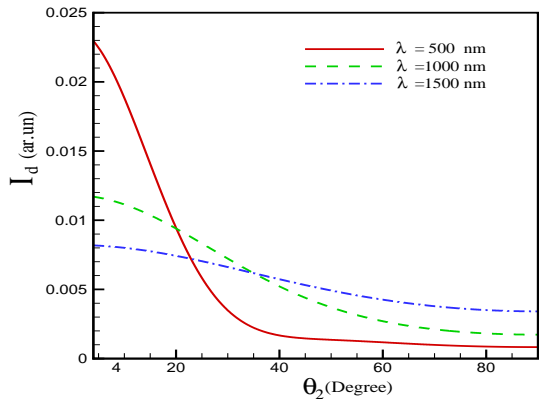}
\caption{Dependence of the diffuse scattered field intensity $I_{d}$
on the angle of scattering $\theta_{2}$, for $H=0.7$,
$\theta_{3}=0$, vertical incident and for different values of
wavelength ($\lambda=500, 1000$ and $1500$ $nm$).}\label{Fig3}
\end{figure}

Two characteristic scales of the problem are roughness and wavelength and in
fact scattering could be modeled by $k\sigma$. When $H$ is constant for a surface,
by increasing $\sigma$ (or increasing $\xi$), diffuse intensity increases for
the specular angle. In other words, increasing $\sigma$ is equivalent to
decreasing the wavelength and because the wavelength is smaller than the
distance between two peaks or correlation length of the surface, reflection
or scattering is larger for the specular angle. But for larger wavelengths
(which are equivalent to smaller $\sigma$ or $\xi$), the wavelength is
larger than (or of the same order as) the distance between two peaks and
the intensity is of the same order for all angles.

\section{Conclusion}

In this paper, we proposed a new perspective to the problem of
scattering from a rough surface in an integer space. We have shown
how this scattering problem is equivalent to reflection phenomenon
in fractional space. In the integer space, fluctuations of the
surface are the main reasons for scattering while in the fractional
space these fluctuations are compensated by the topology of the
problem. To clarify our perspective, we have considered the
Kirchhoff theory of scattering both in the integer dimensional and
fractional spaces and we have shown that the two approaches are
comparable.

\end{document}